\documentstyle[12pt,epsf,epsfig]{article}
\begin{document}
\begin{titlepage}

\pagenumbering{arabic}


\vspace*{1.0cm}
\begin{center}
\Huge {\bf The hypothesis of a real Cabibbo Kobayashi Maskawa matrix }\\
\vspace*{0.8cm}
\centerline{\large Preliminary}	
\vspace*{1.0cm}
\large {P. Checchia \footnote{I.N.F.N., Padova (Italy)} ,
 E. Piotto \footnote{University of Milano, Milano (Italy)}, 
F. Simonetto \footnote{University of Padova, Padova (Italy)} \\}
\vspace*{0.6cm}
\end{center}

\vfill\vfill

\vspace{\fill}

\end{titlepage}

\newpage
\vspace{\fill}

\pagebreak
\section{ Introduction}

In the frame of the Standard Model the mass eigenstates of the quark fields  
are not eigenstates of the weak interactions: the two set of bases are 
connected by a unitary transformation which is represented by a three
times three unitary matrix. This fact accounts for flavour changing charged 
currents  and flavour changing neutral currents 
The mechanism of flavour mixing was originally proposed to account for the different
amplitudes of the decays of the muon and of the  down and strange quarks,
all mediated by weak charged currents \cite{ref:cabi}. It was extended to three quark 
generations in 1973 \cite{ref:koba}
(the beauty quark, down type member of the third generation was later
discovered in 1977) as a possible description 
of the CP violation in $K^0$ decays 
within the frame of the Standard Model without introducing new (so called
Super Weak) interactions. Once removed all arbitrary phases, the three by three unitary
matrix must contain four independent parameters, three rotation angles in the quark field space 
and a phase, which introduces an imaginary part in the Hamiltonian. Therefore the amplitudes for a
process and for its CP conjugated (obtained by hermitian conjugation) may differ.

Several representations of the 
the  Cabibbo-Kobayashi-Maskawa  \cite{ref:cabi,ref:koba} mixing matrix
\begin{equation}
V_{CKM} = \left(
\begin{array}{ccc}
V_{ud} & V_{us} & V_{ub} \\
V_{cd} & V_{cs} & V_{cb} \\
V_{td} & V_{ts} & V_{tb} \end{array}
\right)
\end{equation}
were proposed in the past. 
The Wolfenstein parametrisation  \cite{ref:wolf} is adopted in this paper
as it naturally describes the measured hierarchy among the parameters. The matrix elements
are expressed in terms of the four parameters $\lambda,~A,~\rho$ and $\eta$
\begin{equation}
V_{CKM} = \left(
\begin{array}{ccc}
1 - \lambda^{2}/2 & \lambda & A\lambda^{3}(\rho - i\eta) \\
-\lambda & 1 - \lambda^{2}/2  & A\lambda^{2} \\
A\lambda^{3}(1 - \rho - i\eta) & -A\lambda^{2} & 1 \end{array}
\right) + {\cal{O}}(\lambda^{4}).
\end{equation}

A simple extension valid to $ {\cal{O}}(\lambda^{6})$ is described e.g. in \cite{ref:paga}. 
The parameter $\eta$ is the complex phase accounting for CP violation
of the weak interactions. The $\lambda$ and $A$ parameters are known with
a good accuracy ($\sim 1 \%$ and $\sim 4 \%$, respectively), while  many contributions 
to extract $\rho$ and $\eta$ from the available measurements exist in the 
literature  \cite{ref:paga,ref:paro,ref:mele}.
To this purpose, the measurements of the CP violation parameter in neutral Kaon decay 
$\left| \epsilon_k \right|$, of the difference between the mass eigenstates in
the $B^0_d - \bar{B^0_d}$ system $\Delta m_d$
and of the ratio $\left| \frac{V_{ub}}{V_{cb}} \right|$ and the lower limit on the 
  difference between the mass eigenstates in
the $B^0_s- \bar{B^0_s}$ system $\Delta m_s$
can be used.
On the other hand, since the only direct experimental evidence for CP violation
is given by  the fact that $\left| \epsilon_k \right| \neq 0$ 
and the effect could be explained in term of models proposed in
 alternative to the Standard Model (see, for instance,   \cite{ref:barb,ref:glas}
and references therein),
it is suggested to remove
the constraint coming from the neutral Kaon system and to investigate the   
results on parameter $\eta$
or to test the hypothesis of a real  $V_{CKM}$ matrix.
In \cite{ref:mele}  two different procedures have been exploited.
In the first one the two parameters $\rho$ and $\eta$ has been fitted to
the experimental values of all available  constraints described therein,
apart that 
coming from measurement of neutral kaon mixing.
In the second one the same fit has been computed but forcing $\eta$ to
zero.
The results of the two procedures in \cite{ref:mele} are opposite.
In this letter the second procedure has been followed but with a different
statistical approach, the Best Linear Unbiased Estimator \cite{ref:blue},
and the conclusion is opposite to the  second
procedure in \cite{ref:mele} but in agreement with the first one.
In this letter, the inclusion
of different data sets is also discussed and the corresponding  results are presented.

\section {Measurements and constraints on $V_{CKM}$ parameters}

The $\lambda$ parameter is the sine of the Cabibbo angle 
\cite{ref:pdg}:
\begin{equation}
 \lambda =\left| V_{us} \right| =sin \theta_c = 0.2196 \pm 0.0023.
\end{equation}

The A parameter depends on the matrix element $\left| V_{cb} \right| = 0.0395 \pm 0.0017$
(obtained from semileptonic decays of B hadrons \cite{ref:pdg}) and on $\lambda$:

\begin{equation}
 A =\frac{\left| V_{cb} \right|}{\lambda^2} = 0.819 \pm 0.035.
\end{equation}

In order to constrain the parameters $\rho$ and $\eta$ without considering 
CP violation 
in the neutral Kaon system, three experimental input are used:

\subsection{ $B^0_d$ oscillations.}

 The mass difference $\Delta m_d$ between the mass eigenstates in the  
 $B^0_d - \bar{B^0_d}$ system has been measured with high precision
\cite{ref:pdg,ref:bosc,ref:bosc99}.
 In the Standard Model it can be related to the CKM parameters 
in the following way:

\begin{equation}
 \left[ (1-\rho)^2+\eta^2 \right] =
\frac{ \Delta m_d} 
 {\frac{G^2_F}{6\pi^2}m^2_t m_{B^0_d} 
 \left( f_{B_d} \sqrt{B_{B_d}}\right)^2 \eta_B F(z) A^2 \lambda^6 }
\end{equation}
where  $m_t$ is the
 the top pole mass scaled according to 
\cite{ref:buras} and $z=m^2_t/m^2_W$. The function $F(z)$ is given by:
\begin{equation}
    F(z) = \frac{1}{4} + \frac{9}{4(1-z)} - \frac{3}{2(1-z)^2} -
        \frac{3z^{2}\ln{z}} {2(1-z)^3}.
\end{equation}
The values of all the parameters are given in {\it table} \ref{tab:param}. 

In the  $\rho - \eta$ plane, the measurement of $\Delta m_d$ corresponds
to a circumference centred in (1,0). By constraining $\eta$  to zero,
an evaluation of $\rho$ can  be obtained. Unfortunately
the term $f_{Bd} \sqrt{B_{B_d}}$, given by lattice QCD calculations, 
is known with a  $20\%$ order  uncertainty  \cite{ref:fbd97,ref:fbd98} 
and  therefore  it gives the largest contribution to the error on the $\rho$
determination.

\begin{table}[htb]
\centering
\begin{tabular}{|c|c|c|} \hline
parameter    & value used in \protect\cite{ref:mele}      & new input \\
\hline
 $G_F	$    &$1.16639(1)\times 10^{-5} \rm{GeV}^{-2}   $  &	     \\
 $\lambda$   &$ 0.2196 \pm 0.0023  $                       &            \\
 $A$         &$ 0.819 \pm 0.035    $                       &            \\
 $m_t$       &$ 166.8 \pm 5.3      $ GeV                   &            \\
 $m_W$       &$ 80.375 \pm 0.064   $ GeV                   &            \\
 $m_{B_d}$   &$ 5.2792 \pm 0.0018  $ GeV                   &            \\
 $m_{B_s}$   &$ 5.3692 \pm 0.0020  $ GeV                   &            \\
 $\eta_B$    &$ 0.55  \pm 0.01     $                       &            \\
$\Delta m_d$ &$ 0.471 \pm 0.016 ~\rm{ps}^{-1} $      &            \\
$\Delta m_s$ &$ > 12.4 ~\rm{ps}^{-1}(95 \% ~\rm{C.L.}) $ & $>12.3 ~\rm{ps}^{-1}(95 \% ~\rm{C.L.})$\cite{ref:bosc99}  \\
$ \left|V_{ub} \right|/\left|V_{cb} \right|$&$0.093 \pm 0.016$ &   $0.090 \pm 0.012$  \cite{ref:cleo99,ref:lepvub}       \\
$f_{B_d} \sqrt{B_{B_d}}$ &$ 0.201 \pm 0.042$ GeV          
                         &$ 0.215^{+0.040}_{-0.030}$GeV \cite{ref:fbd98},$~0.210 ^{+0.039}_{-0.032}$ GeV \protect\cite{ref:paro99}  \\
$ \xi$       &$ 1.14 \pm 0.08 $ &$ 1.14^{+0.07}_{-0.06} $ \protect\cite{ref:fbd98}     \\
\hline
\end{tabular}
\caption{ Physical parameters used in the formulae (5),(6),(7) and (8).
In the second column the values used in \protect\cite{ref:mele} and
in the third column  the most recent values
are given.
}
\label{tab:param}
\end{table}

\subsection{ $B^0_s$ oscillations.}

The mass difference $\Delta m_s$ between the mass eigenstates in the  
 $B^0_s - \bar{B^0_s}$ system  is expected 
to be much larger than  $\Delta m_d$ and in the Standard Model is related to
the CKM parameters:
\begin{equation}
 \left[ (1-\rho)^2+\eta^2 \right] =
 \frac{\Delta m_d} { \Delta m_s} \frac{1}{\lambda^2} \frac{ m_{B^0_s}}{ m_{B^0_d}} 
 \xi^2
\end{equation}

Since the ratio 
\begin{equation}
 \xi = \frac{  f_{B_s} \sqrt{B_{B_s}}} {f_{B_d} \sqrt{B_{B_d}}}
\end{equation}
is computed by lattice QCD with a better precision than the single terms,
a measurement of $\frac{ \Delta m_d}{ \Delta m_s} $ could provide a much 
stronger constraint on the $\rho - \eta$ plane. However, given the very high 
frequency in the  $B^0_s - \bar{B^0_s}$ system oscillation, 
only a lower limit on  $ \Delta m_s$ is available, 
as shown in {\it table} \ref{tab:param}, which  corresponds to a circular
bound  in the two parameter space or, if the assumption $\eta = 0$ is
made, to a lower limit for $\rho$.

\subsection {$ \left|V_{ub} \right|$ measurements from semileptonic b decay.}
 
Charmless semileptonic b decays have been used to measure $ \left|V_{ub} \right|$ 
or the ratio $ \left|V_{ub} \right| / \left|V_{cb} \right|$.
The CLEO collaboration determined that parameter both by measuring the rate 
of leptons produced in B semileptonic decays beyond the charm end-point \cite{ref:cleo} 
and
from direct reconstruction of charmless B semileptonic decay \cite{ref:cleo2}. 
The two results are consistent
but both methods are limited by theoretical uncertainties. In \cite{ref:pdg} 
they are not combined and a 
value $\left|V_{ub} \right|/\left|V_{cb} \right|=0.08\pm0.02 $
obtained from the former result is given.
Recently CLEO gave a new result and an average with their previous 
results \cite{ref:cleo99}.
At LEP, ALEPH \cite{ref:alvub},
 L3 \cite{ref:l3vub} and
DELPHI  \cite{ref:devub} have measured the inclusive charmless semileptonic
transitions \mbox{$b \rightarrow ul\nu$}. The average $\left|V_{ub} \right|$
value from LEP measurements 
is given in \cite{ref:lepvub} 
and the combination obtained  using the latest CLEO result   
is given in  {\it table} \ref{tab:param} in terms 
of $\left|V_{ub} \right|/\left|V_{cb} \right|$. 
The ratio of the two CKM matrix elements is 
related to the $\rho$ and $\eta$ parameters by:

\begin{equation}
 \frac {\left|V_{ub} \right| }{ \left|V_{cb} \right|}= \lambda \sqrt{\rho^2 + \eta^2}
\end{equation}
and hence, in the  $\rho - \eta$ plane, 
the measurement of  $ \left|V_{ub} \right| / \left|V_{cb} \right|$ corresponds
to a circumference centred at the origin. If $\eta$ is assumed to be zero, it would
be proportional to  the $\rho$ absolute value.

\section {Data compatibility with a real CKM matrix hypothesis}

The assumption of a real CKM matrix implies that all the constraints described
in the previous section are reduced to values of (or limits on) $\rho$.
The compatibility of the obtained values can then be used to estimate the goodness of the
assumption itself.
In  \cite{ref:mele} and in the explicit reference to it in  \cite{ref:glas},
it is written that the hypothesis of a real CKM matrix can fit the data.
However it is unclear 
which was the  statistical approach followed
in the second procedure in \cite{ref:mele} to come to that statement
%
\footnote{In the cited paper, a first procedure where both $\rho$ and $\eta$ 
are fitted 
excludes the hypothesis $\eta=0$ at the $99 \%$ C.L.. A second  fit
with $\eta$ forced to zero gives  $\chi^2=6.7$ and the author concludes that 
this is compatible with a real CKM hypothesis
but it is not
specified why the mentioned $\chi^2$ value corresponds to a reasonable Confidence Level
for such hypothesis.}
%
and a completely
different conclusion can be obtained with the same procedure and a
standard statistical method.
In addition the claim in the second procedure of \cite{ref:mele}, 
although using a different input-data-set,
contradicts what is reported in \cite{ref:barb}. 
Assuming $\eta=0$ , modifying accordingly eq. (5), (7) and (9) and using 
exactly the same input parameters as \cite{ref:mele} 
({\it table} \ref{tab:param} second column), the values

\begin{equation}
 \rho^{\Delta {m_d}}=0.01^{+0.18}_{-0.26}
\end{equation}
and
\begin{equation}
 \rho^{V_{ub}}=\pm (0.42 \pm 0.07)
\end{equation}

are obtained from eq. (5) and 
(9), respectively.
The limit on $\Delta m_s$  has been obtained by means of the
amplitude method \cite{ref:moser}
which allows to know the exclusion Confidence
Level  for any value of  $\Delta {m_s}$.
Therefore, by a convolution with the dominant uncertainty from the ratio $\xi$ in eq. (7)
it is possible to obtain: 
\begin{equation}
 \rho^{\Delta m_s}> -0.05 
\end{equation}
at the $95 \%$ Confidence Level.  

In a naive approach on which the errors on $\rho^{\Delta {m_d}}$ and $ \rho^{V_{ub}}$
are assumed to be uncorrelated
it is evident that the two values are fairly incompatible.
The negative $ \rho^{V_{ub}}$ solution is clearly excluded by the 
 $\rho^{\Delta m_s}$ limit and therefore it can be discarded.
In order to include all the correlations 
due to common terms contributing to the errors, namely $\left| V_{cb}\right|$ and $\lambda$,
a two dimensional error matrix $\bf M$ has been written and the 
Best Linear Unbiased Estimator
\cite{ref:blue} has been used:
\begin{equation}
 \rho^{BLUE}=\frac {\Sigma^2_{i=1}\Sigma^2_{j=1} \rho_i({\bf M}^{-1})_{ij} }
                   {\Sigma^2_{i=1}\Sigma^2_{j=1}       ({\bf M}^{-1})_{ij} }
\end{equation}  
with the variance
\begin{equation}
 \sigma^2_{\rho}=\frac {1 }
                   {\Sigma^2_{i=1}\Sigma^2_{j=1}       ({\bf M}^{-1})_{ij} }.
\end{equation}  
The error matrix $\bf M$ includes correlated and uncorrelated contributions:
\begin{equation}
M_{ij}=\delta_{ij}\sigma^{uncorr}_i\sigma^{uncorr}_j + \Sigma_{\alpha=1}^m \Delta_{\alpha i} \Delta_{\alpha j}
\end{equation}  
where the indexes $i$ and $j$ run over the two $\rho$ measurements and 
$\Delta_{\alpha i}$ is the change (with sign) on measurement $i$ when the common
systematic parameter $\alpha$ is moved by its error. 
The $\chi^2$ 
is  obtained by:
\begin{equation}
 \chi^2=
 \Sigma^2_{i=1}\Sigma^2_{j=1} \left( 
 \left[ \rho_i-\rho^{BLUE}\right] ({\bf M}^{-1})_{ij} \left[ \rho_j-\rho^{BLUE}\right]
 \right). 
\end{equation}

With the 
{\it table} \ref{tab:param} (second column)  parameters and taking the positive error 
in eq. (10),
\begin{equation}
   \rho^{BLUE}=0.36 \pm 0.07  ~ \rm{and} ~ \chi^2=4.3~ \rm{(1~ degree~ of ~freedom)}
\end{equation}
are obtained. 
The values $\Delta_{\alpha i}$ 
and 
the more relevant uncorrelated contributions
are given in   {\it table} \ref{tab:delta} and   
{\it table} \ref{tab:uncor} respectively. 
The correlation between the two measurements
is small given the dominance of the uncertainty on  $f_{B_d} \sqrt{B_{B_d}}$ 
in $\rho^{\Delta m_d}$ as it can be deduced from the small off-diagonal term in the
error matrix $\bf M$ ({\it table} \ref{tab:M}). 
The corresponding  $\chi^2$ probability is $3.9 \%$ and this is 
clearly in contradiction with what stated after the second procedure of \cite{ref:mele}. 

\begin{table}[htb]
\centering
\begin{tabular}{|c|c|c|} \hline
parameter ($\alpha$) & $\Delta_{\alpha  \rho^{\Delta m_s}}$ &  $\Delta_{\alpha  \rho^{V_{ub}}}$\\
\hline
 $V_{cb} $   &  4.3     & -1.8	     \\
 $\lambda$   &  1.1     & -0.4       \\
\hline
\end{tabular}
\caption{Variation  ($\times 10^{-2}$)
 on $\rho$  when the parameter $\alpha$ is moved by its positive error.
}
\label{tab:delta}
\end{table}

\begin{table}[htb]
\centering
\begin{tabular}{|c|c|c|} \hline
parameter               & $ i= \rho^{\Delta m_d}$ & $ i=\rho^{V_{ub}}$\\
\hline
 $f_{B_d} \sqrt{B_{B_d}}$ &$~+17.1~-26.1$&             \\
 $m_t$                    & $\pm3.1$&             \\
$\Delta m_d$              & $\pm1.7$&             \\           
 $\eta_B$                 & $\pm$1.0&             \\
$ \left|V_{ub} \right|$   &    & $\pm 7.1$        \\
\hline
\end{tabular}
\caption{Uncorrelated contributions ($\times 10^{-2}$) to the $\rho$ errors
$\sigma_i$ computed from the parameters listed in
 the second column of {\it table} \ref{tab:param}. 
}
\label{tab:uncor}
\end{table}
\begin{table}[htb]
\centering
\begin{tabular}{|c|c|c|} \hline
             & $j=\rho^{\Delta m_d}$ & $ j=\rho^{V_{ub}}$       \\
\hline
$i=\rho^{\Delta m_d}$ & $3.3 \times 10^{-2}$ &$ -0.8 \times 10^{-3}$ \\
$i=\rho^{V_{ub}}$    & $-0.8  \times 10^{-3}$&$5.4 \times 10^{-3}$   \\
\hline
\end{tabular}
\caption{ Elements of the error matrix $\bf M$ 
 $~M_{ij}$ computed from the parameters listed in
the second column of {\it table} \ref{tab:param}. }
\label{tab:M}
\end{table}

The effect
of the limit on  $\Delta {m_s}$ is negligible even though the information 
contained in the amplitude is taken into account as suggested in \cite{ref:paga}:
since the value of $\Delta {m_s}$ that one would obtain by inserting 
$\rho= \rho^{BLUE}$ and 
$\eta=0$ in eq. (7) is $\Delta {m_s}=31~ \rm{ps}^{-1} $, the present 
experimental sensitivity ($13.8~ \rm{ps}^{-1} $)
 does not give any sizeable information 
on that $\rho$ region.

In order to take into account terms of the order up to 
$O(\lambda^5)$
 \cite{ref:buras2}, the substitution
  $\rho \rightarrow \bar{\rho}=\rho (1-\lambda^2/2) $
must be  done in eq. (5) 
with $\eta=0$.

More recent values of $f_{B_d} \sqrt{B_{B_d}}$ can be  taken from 
\cite{ref:fbd98} (option a: $f_{B_d} \sqrt{B_{B_d}}= 0.215^{+0.040}_{-0.030}$ GeV) or  from 
\cite{ref:paro99}  (option b: $f_{B_d} \sqrt{B_{B_d}}= 0.210^{+0.039}_{-0.032}$ GeV) 
where the latter value has been obtained by considering the 
measured value of  $f_{D_s}$ and the theoretical
evaluation of the ratio $f_{B_d}/f_{D_s}$. In the following both cases will be considered.
Including the last results on $V_{ub}$ \cite{ref:cleo99,ref:lepvub} ( see {\it table} \ref{tab:param}),
the incompatibility of the two $\bar{\rho}$ measurements is still present: 
\begin{equation}
   \bar{\rho}^{BLUE}=0.36 \pm 0.05
   ~\rm{and}~\chi^2=3.9~ {\rm(1~ degree~ of~ freedom)}
\end{equation}
corresponding to a  $\chi^2$ probability of $4.9 \%$ (option a) and 
\begin{equation}
   \bar{\rho}^{BLUE}=0.35 \pm 0.05
   ~\rm{and}~\chi^2=4.3~ {\rm(1~ degree~ of~ freedom)}
\end{equation}
corresponding to a  $\chi^2$ probability of $3.7 \%$ (option b).  

Since the dominant error 
in $\rho^{\Delta m_d}$ is due to the lattice QCD computation, the hypothesis of
a flat error with the same R.M.S. on $f_{B_d} \sqrt{B_{B_d}}$ has been studied 
with a simple simulation. Several experimental results for 
$\rho^{\Delta m_d}$ and $\rho^{V_{ub}}$
have been generated 
with central values equal to $\rho^{BLUE}$. For $\rho^{\Delta m_d}$  this is achieved
by shifting the value of $f_{B_d} \sqrt{B_{B_d}}$ and allowing it to vary within
a flat distribution  with the R.M.S. corresponding to the quoted error. All the other 
parameters of eq. (5)
are allowed to vary with a gaussian distribution
corresponding to their error. In each experiment the combined value and the
$\chi^2$ are computed according to eq. (13) and (16), respectively.
 Looking at the $\chi^2$ 
 distribution, it is possible
to determine the fraction of simulated experiments with a 
$\chi^2$ higher than the
value found in the evaluation with real data.
This procedure has been repeated for the data sets of {\it table} \ref{tab:param}
and for none of them the $\chi^2$ has been found to be higher than the
experimental values of eq. (17), (18) and (19) in more than $5\%$ of the cases.

In \cite{ref:barb} it is mentioned that contributions from
new physics can modify the value 
$\rho^{\Delta m_d}$ and the limit $\rho^{\Delta m_s}$.
In particular, being $\delta_{d,s}$ the new physics contribution to 
$\Delta m_{d,s}$,  the changes on $\rho^{\Delta m_{d,s}}$ depend on the
fractional contributions $F_{d,s}=\delta_{d,s}/\Delta m_{d,s}$ and with 
the data  of the third column of {\it table} \ref{tab:param}
$\Delta\rho^{\Delta m_d}\sim 0.5F_d$ and 
$\Delta\rho^{\Delta m_s}\sim 0.5(F_d-F_s)$ are obtained.
It is straightforward that in presence  of a very large term $\Delta\rho$
there cannot be any exclusion of the $\eta=0$ hypothesis. It must be noticed,
 however, that those terms were not foreseen in the second procedure of \cite{ref:mele} 
which has then to be compared with the present result from eq. (17). 
An evaluation of the allowed range of $F_d$ 
from the data of {\it table} \ref{tab:param} $3^{rd}$ column option b), shows that  
small positive values of $F_d$ are excluded 
($F_d>0.04$ at the $95\% $ Confidence Level) while all  negative values
are excluded except for  the case of large
positive $F_s$  (${\cal{O}}(30\%)$)
 which would allow for negative values of $\rho$. This evaluation agrees with 
\cite{ref:barb}. 

\section {A method to determine $ \bar{\rho}$ and $\bar{\eta}$}

To obtain $ \bar{\rho}$ and $\bar{\eta}$ without using the constraint on 
$\left| \epsilon_k \right|$ the eq. (5), (7) and (9) have to be used. 
It must be noticed that, in absence of a measurement of $\Delta m_s$,
the number  of constraints is such that a solution for the system
of eq. (5) and eq. (9) is often found or, in other words, a 0(C) fit
with $ \bar{\rho}$ and $\bar{\eta}$ as free parameters could converge 
to a pair of values such that the total $\chi^2$  function
minimum is zero. The addition of terms taking into account 
the existing uncertitude in the value of the parameters entering into
the two expressions does not modify the number of degrees of freedom. 
With the present data ({\it table} \ref{tab:param} $3^{rd}$ column option b)
such a solution exists and it is:
\begin{equation}
   \bar{\rho}=0.13_{-0.23}^{+0.13}  ~ \rm{and} ~ \bar{\eta}=0.38_{-0.09}^{+0.07}.
\end{equation}

In the most recent search for $B^0_s$ oscillation \cite{ref:bosc99}
with the amplitude method
an evidence for such a signal was not found but in 
the region 13 ps$^{-1} < \Delta m_s< 17 \rm{ps}^{-1}$
the combined amplitude was greater than zero by more than
1.645 standard deviations. As explained in  \cite{ref:moser}, 
the amplitude $A$  and its error
$\sigma_{A}$
at any value of $\Delta m_s$
can be related to the log-likelihood referenced to its
value for an infinite $\Delta m_s$ (infinite oscillation
frequency) by the expression 
$\Delta {\cal{L}}^{\infty}(\Delta m_s) =[ 1/2- A(\Delta m_s)]/\sigma_{A(\Delta m_s)}^2$.
From eq. (7) is possible to express $\Delta m_s$ as function of $\bar{\rho},\bar{\eta},\xi$ etc.
and then the experimental amplitude and its error corresponding to a set of these parameter can be determined.  
Therefore a global log-likelihood function
 can be  written 
as: ${\cal{L}}= \Delta {\cal{L}}^{\infty}( \bar{\rho},\bar{\eta},\xi)
    +{\cal{L}}^{\Delta m_d , V_{ub}}(\bar{\rho},\bar{\eta}, f_{B_d} \sqrt{B_{B_d}},..)$
where ${\cal{L}}^{\Delta m_d , V_{ub}}$ is the log-likelihood
obtained from eq. (5) and (9) as function of the input parameters of those equations and their errors.
Clearly ${\cal{L}}^{\Delta m_d , V_{ub}}$ alone has a minimum in correspondence
to the values of eq. (20) while the minimisation of ${\cal{L}}$ gives:
\begin{equation}
   \bar{\rho}=0.14_{-0.06}^{+0.05}  ~ \rm{and} ~ \bar{\eta}=0.37 \pm 0.05.
\end{equation}
The addition of the term $\Delta {\cal{L}}^{\infty}$ has the effect of excluding 
the $\bar{\rho},\bar{\eta}$ space on the region corresponding to the excluded
$\Delta m_s$ values (i.e. negative $\bar{\rho}$)
 but also of reducing considerably the errors  of the two
parameters on the opposite direction. The latter fact is related to the 
presence  of a minimum on $\Delta {\cal{L}}^{\infty}(\Delta m_s)$ for
$\Delta m_s=14.75~\rm{ps}^{-1}$ (see \cite{ref:bosc99}) which is about 2.8 units
below the asymptotic value for very high frequencies \cite{ref:ampl99}.
However it has to be emphasised that, in absence of a clear signal for 
the $B^0_s$ oscillation, the presence of ghost minima on  $\Delta {\cal{L}}^{\infty}(\Delta m_s)$ 
cannot be excluded. Moreover, the minimum is slightly above the experimental sensitivity 
$ (14.3~\rm{ps}^{-1}$). As a consequence, the results of eq. (21) must be taken 
as an indication of a  method to include the $\Delta m_s$ information 
(and of the relevant effects of its inclusion)   
rather than as a robust evaluation of  
$\bar{\rho}$ and $\bar{\eta}$ with the present data. 

\section{Conclusions}

The hypothesis of a real CKM matrix is tested on the basis of the present published
and preliminary data and lattice QCD calculations. The Best Linear Unbiased Estimator  
has been used for the statistical approach. 
With all the input data used,
included those suggested in \cite{ref:mele}, that hypothesis is excluded at more than
$95 \%$ Confidence Level. This result 
agrees with the result indicated in \cite{ref:barb} with an older input data set
and with a  first procedure reported in   \cite{ref:mele} 
but contradicts statements in
the same paper from a second procedure.
A method to include the $\Delta m_s$  information
is proposed showing that  a strong reduction on the error
on $\bar{\rho}$ and $\bar{\eta}$ is achievable.
\section{Acknowledgements}
We wish to thank R. Barbieri, M. Loreti, G. Martinelli, M. Mazzucato, F.Parodi, A. Stocchi
and L. Ventura for 
comments and useful discussions.



\end{document}